\documentclass[useAMS,usenatbib]{mn2e}

\usepackage{graphicx}
\usepackage{amsmath}
\usepackage{amssymb} 
\usepackage{longtable} 
\title[Globular Clusters in Sombrero]{
Spectra of globular clusters in the Sombrero galaxy: evidence for spectroscopic metallicity
bimodality\thanks{Based on observations obtained at the W. M. Keck
Observatory, which is operated as a
scientific partnership among the
California Institute of Technology, the University of California, and the
National Aeronautics and Space Administration.}\\}
\author[A. Alves-Brito et al. 2011]{Alan Alves-Brito$^{1,2}$\thanks{E-mail:
abrito@astro.puc.cl (AAB)}, George K.T. Hau$^{1,3}$, Duncan A. Forbes$^{1}$, 
Lee R. Spitler$^{1}$
 \\\\
\normalfont{\LARGE{Jay Strader$^{4}$, Jean P. Brodie$^{5}$, Katherine L. Rhode$^{6}$}}\\
$^{1}$ Centre for Astrophysics and Supercomputing, Swinburne University of
Technology, Hawthorn, Victoria 3122, Australia\\
$^{2}$ PUC, Departamento de Astronomia y Astrofisica, Av. Vicuna Mackenna 4860, Santiago, Chile\\
$^{3}$ European Southern Observatory, Alonso de Cordova 3107, Vitacura,  Santiago, Chile\\
$^{4}$ Harvard-Smithsonian Centre for Astrophysics, 60, Garden St., Cambridge, MA 02144, USA\\
$^{5}$ UCO/Lick Observatory, University of California, Santa Cruz, CA 95064,
USA\\
$^{6}$ Department of Astronomy, Indiana University, Bloomington, IN 47405, USA\\
}
\begin{document}

\date{Accepted 1988 December 15. Received 1988 December 14; in original form 1988 October 11}

\pagerange{\pageref{firstpage}--\pageref{lastpage}} \pubyear{2002}

\maketitle

\label{firstpage}

\begin{abstract}

We present a large sample of over 200 integrated-light spectra of confirmed globular
clusters (GCs) associated with the Sombrero (M104) galaxy taken with the DEIMOS instrument on the Keck telescope. 
A significant fraction of the spectra have signal-to-noise levels high enough to allow measurements of GC metallicities using the method of Brodie \& Huchra (1990). We find a distribution of spectroscopic metallicities ranging from --2.2 $<$ [Fe/H] $<$ +0.1 
that is bimodal, with peaks at [Fe/H] $\sim$ --1.4 and --0.6. Thus the GC system of the Sombrero galaxy, like a few other galaxies now studied in detail, reveals a bimodal {\it spectroscopic} metallicity distribution supporting the long-held belief that colour bimodality reflects two 
metallicity subpopulations. This further suggests that the transformation from
optical colour to metallicity for old stellar populations, such as GCs, is not
strongly non-linear. We also explore the radial and magnitude distribution with
metallicity for GC subpopulations but small number statistics prevent any clear trends in these
distributions.

\end{abstract}
 
\begin{keywords}
globular clusters: general, galaxies: individual: Sombrero (M104: NGC 4594), galaxies: star clusters.
\end{keywords}

\section{Introduction}

Understanding how galaxies were formed and have evolved remains one of
the basic problems to be solved in astrophysics. In this context, extragalactic
globular clusters (GCs) play an important role because their formation is
associated with the physical processes occurring before galaxies were
entirely assembled (see Brodie \& Strader 2006 for a review).

It has been recognised for some time that massive galaxies have bimodal GC colour distributions (Ashman \& Zepf 1993). 
This suggests two modes, or phases, of GC formation. Indeed, several models have been proposed to explain the different modes in terms 
of variations in the epoch of GC formation, stellar population properties and the timescale of galaxy assembly (Ashman \& Zepf 1992; 
Zepf \& Ashman 1993; Forbes et al. 1997; Cote et al. 1998; Beasley et al. 2002). As most GCs appear to be very old (see review by Brodie \& Strader 2006), the bimodality in colour is 
normally assumed to translate directly into a bimodality in metallicity. 
However, doubt has recently been cast on this interpretation by some (e.g. Yoon et al.
(2006; Blakeslee, Cantiello \& Peng 2010). Depending on the degree to which
the transformation of colour into metallicity is non-linear, it is possible for
an instrinsically unimodal metallicity distribution to appear bimodal in colour
space. If correct, this would have direct implications for the observed blue
tilt (a trend for redder colours with higher luminosities in the blue GC
subpopulation) and the correlation of the mean GC colour with galaxy luminosity. It
would also radically change current ideas for two modes of GC formation.

As noted by Blakeslee et al. (2010), {\it "Very little is actually known about
the detailed metallicity distributions of GCs in giant ellipticals."}.  Two
giant ellipticals with optical 
spectroscopic metallicities derived for their GC systems
include M49 with 47 measurements (Strader et al. 2007) and NGC 5128 with over
200 GC spectroscopic metallicities (Beasley et al. 2008). 
In both cases the distribution of GC spectroscopic metallicities are
bimodal. The GC systems of these galaxies also have
bimodal optical colour distributions, as do most but not all galaxies (see e.g. Foster et al. 2010).
Interestingly, the infrared study of GCs in NGC 5128 by Spitler et al. (2008),
which is more sensitive to metallicity than optical colours, also revealed
colour bimodality in the 146 GCs they studied. Along similar lines, Kundu \& Zepf (2007) used optical-infrared colours of 80 GCs in M87 to show their distribution was bimodal. 
Thus these giant ellipticals with optical spectra and/or infrared colours of GCs show metallicity bimodality. 

Recently attempts have been made to derive spectroscopic metallicities using the
infrared Calcium Triplet (CaT) lines (at 8498, 8542, and 8662 $\rm \AA$)
for the GC systems of two giant ellipticals (Foster et al. 2010, 2011).
In the case of NGC 1407, the CaT metallicity distribution of 144 GCs is better described as 
unimodal, whereas Cenarro et al. (2007) suggested that the distribution was bimodal based on metallicities from optical wavelength spectra of just 20 GCs. For NGC 4494, the CaT metallicity distribution again appears unimodal. Unfortunately,  there is no published work based on optical wavelength metallicities for the GC system in this galaxy. Clearly a large sample study of GC metallicities derived from both blue and infrared absorption lines is needed to resolve these issues (see Foster et al. 2010 for a complete discussion of potential factors affecting the CaT derived metallicities). In summary, it appears that 
at least {\it some} elliptical galaxies reveal the presence of two metallicity subpopulations in their GC systems but the number of systems studied remains small. 

For late-type spiral galaxies, only two are well-studied (i.e. the Milky Way and
M31) and both reveal bimodal spectroscopic metallicity distributions for their
GC systems (e.g. Zinn 1985; Barmby et al. 2000 but see also Caldwell et
al. 2011). No early-type spirals
have published spectra for large numbers of GCs.


The nearest spiral galaxy with a large GC system is the Sombrero
galaxy (M104, NGC 4594), which lies a distance of 9.0 $\pm$ 0.1 Mpc (see Table 4
of Spitler et al. 2006 for a summary). Although classified as an edge-on Sa galaxy, it has a bulge-to-total ratio of 0.8 (Kent 1988) and hence might  be  better described 
as a massive bulge plus an extended disk. This galaxy hosts some 1900 GCs
that extend to a projected radius of 50 kpc (Rhode \& Zepf
2004; Spitler et al. 2006) and one Ultra Compact Dwarf (Hau et al. 2009).
A number of previous studies have obtained spectra of Sombrero 
GCs (Bridges et al. 1997; Larsen et al. 2002; Held et al. 2003; 
Bridges et al. 2007), however the number of spectra of sufficient 
signal-to-noise (S/N) to measure individual GC metallicities was very
limited. The Keck spectra of Larsen et al. (2002) confirmed that the
dozen luminous GCs they studied were all old. 

In this paper we present Keck integrated spectra for over 200 
globular clusters in M104. This is the largest sample of GCs in the Sombrero galaxy
homogeneously analysed to date. Our main goal is to obtain radial
velocities to confirm association with the Sombrero galaxy and to obtain spectroscopic metallicities. A future paper will 
investigate the kinematic properties of the GC system. 

\section{Observations and data reduction}

GC candidates were selected from two published photometric studies. In the
inner regions we used the $BVR$ HST/ACS mosaic of Spitler et al.~(2006),
which covers approximately the central 10\arcmin $\times$ 7\arcmin ~of the
galaxy with minimal contamination. Beyond this area, we used the
ground-based $BVR$ catalogue of Rhode \& Zepf (2004). This latter catalogue
has higher contamination but extends to the apparent edge of the GC system
at a projected radius of 19\arcmin ~and so is necessary to complement the
wide field-of-view (16\arcmin $\times$ 5\arcmin) of the DEIMOS spectrograph. The central dust lane of NGC 4594 was excluded from our target selection. 

Target selection was made using both magnitude and colour.
The colour selection using the $BVR$ filters was
performed using previously published GCs as a guide
(Bridges et al. 1997; Larsen et al. 2002; Held et al. 2003; 
Bridges et al. 2007). Four masks were
designed, with a total of 404 GC candidates. The median candidate had $V
\sim 22$ with both colour subpopulations well-represented.  The masks
included candidates with a wide radial range (extending to $\sim
11$\arcmin) and a moderate range in position angle, with the NE and SW
quadrants of the galaxy well-covered.

All four masks were observed during a single run from 2006 April 26--29. 
Total exposure times for each mask ranged from 3--4.5 hr, divided into
individual exposures of 30 min. All spectroscopic observations were made
with the 900 l/mm grating, blazed at 5500 \AA. Slits were milled to a
width of 1.0\arcsec, yielding a resolution of 2.1 \AA~ (FWHM). Seeing was
variable during the run, ranging from 0.6--1.1\arcsec. 

\begin{figure}
\includegraphics[width=88mm]{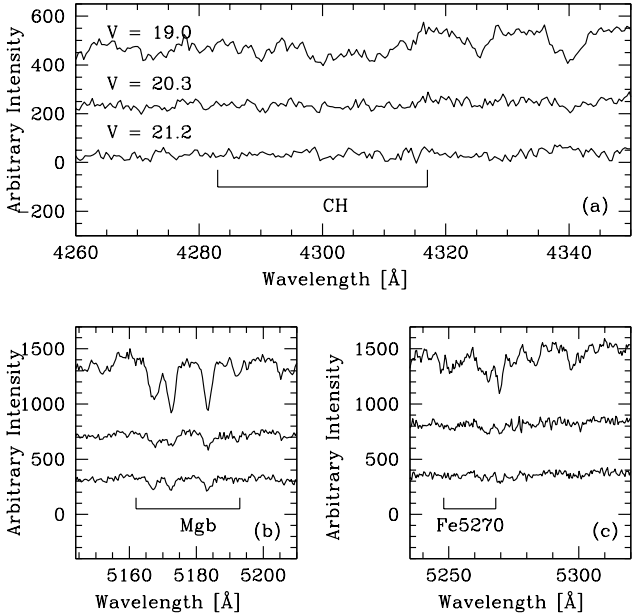}
 \caption{Example GC spectra obtained by the DEIMOS spectrograph. Shown are rest wavelength 
 spectra around the metallicity sensitive indices of CH or G band (a), Mgb (b) and Fe5270 (c)
for different GC magnitudes, as indicated in the figure.}
 \label{f:espectros}
\end{figure}

The DEIMOS data were reduced using the DEEP2 {\tt spec2d} package, adapted for
our instrument setup. The nominal
spectral coverage for an object in the center of the mask was $\sim
3700$--6900 $\AA$, but the throughput toward the blue was low, so the useful
spectral range was somewhat smaller.
The wavelength scales were calibrated using arc spectra and
fitting a 4th degree polynomial to the wavelength solution. The spectra
were divided into two regions (``blue" and "red") and the wavelength solutions
calculated separately, with an RMS scatter of 0.08 $\AA$~ in the blue and 0.06 $\AA$~ in the red,
corresponding to roughly 5 and 3 kms$^{-1}$ respectively. The presence of
bright sky lines in the red spectra also allowed a correction for any
instrument flexture between the science data taken during the night
and the arcs taken during the day.
Examples of the final spectra are given in Fig. \ref{f:espectros}.

\section{Results and discussion}

\subsection{Radial Velocity Measurements}

Heliocentric radial velocities were obtained
using the IRAF\footnote{IRAF is distributed by the National
Optical Astronomy Observatory, which is operated by the Association of
Universities for Research in Astronomy (AURA) under cooperative agreement with
the National Science Foundation.} task {\tt fxcor}.  
{\tt Fxcor} performs Fourier
cross-correlation of input object spectra against a specified template spectrum.
For the cross-correlation of the DEIMOS spectra we used the spectrum of the K
giant star Arcturus as the template. 

We were able to measure radial velocities for 258 GC candidates. 
Data for all of the objects for which we could measure a radial velocity are presented in Table A. 
In this table, columns 1 and 2 indicate the
object's ID (taken from either Spitler et al. 2006 or Rhode \& Zepf 2004, indicated
as S or RZ respectively); column 3 the mask number; 
columns 4--5 give the J2000.0 equatorial coordinates; columns 6--8 give the major- (X), minor-axis
(Y) and the projected distances (R$_{p}$), respectively, from the centre of
the galaxy; columns 9--10 give the radial velocities and their errors. 
The final velocity listed is the mean of the blue and red velocities, 
corrected to heliocentric. The uncertainties quoted are derived from the width
of the cross-correlation peak, normalized using the distribution of the differences between
the blue and red velocities as a function of magnitude, with a minimum value of 8 kms$^{-1}$.
These uncertainties primarily reflect the random component of the error; 
we estimate our systematic error, due primarily to zero-point uncertainties, to be $\le$ 20 kms$^{-1}$. 

To obtain the galactocentric coordinates (X,Y) and projected distances
for our sample we adopted a distance to Sombrero of 9.0~Mpc (Spitler et~al. 2006). 
At this distance, 1 arcmin corresponds to $\sim$ 2.62 kpc. We used a central
position of $\alpha_{\rm J2000}$ = 12$^{\rm h}$39$^{\rm
m}$99$^{\rm s}$.43, $\delta_{\rm J2000}$ = $-$11$^{\rm o}$37'23''. 
The position angle for the X-coordinate is taken as 90$^{\rm o}$ (e.g., de
Vaucouleurs et al. 1991). Hence, we are able to probe 
the Sombrero GC system out to $\sim$ 30 kpc. 

\begin{figure}
\includegraphics[width=88mm]{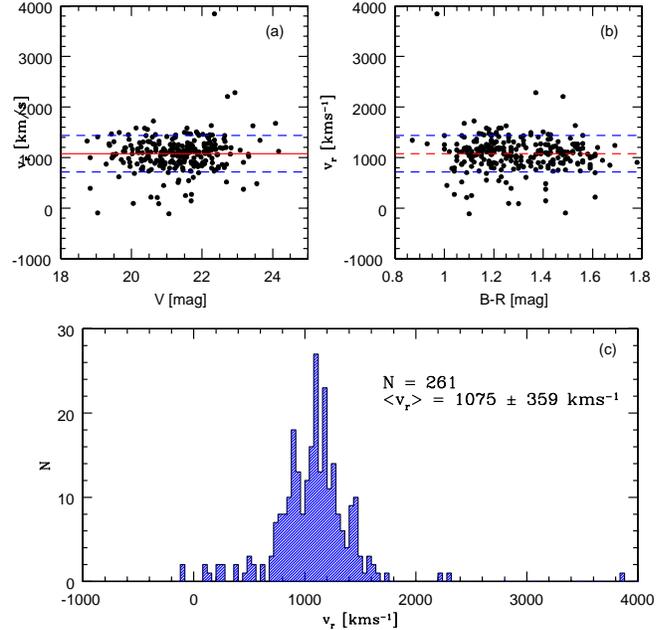}
 \caption{Radial velocities of GC candidates. Velocity as a function of V band magnitude (a) and colour
 (b) are shown, along with the radial velocity distribution (c). 
 The horizontal lines in (a) and (b) mark the mean v$_{r}$ value ({solid
 line}) and the 1$\sigma$ dispersion ({dotted lines}).}
 \label{f:vrmag}
\end{figure}

Figure \ref{f:vrmag} shows the radial velocities as a
function of V magnitude, $B-R$ colour, and the radial velocity
distribution for the 258 GC candidates. The mean velocity of the GC candidate
sample is v$_{r}$ = 1075 $\pm$ 359
kms$^{-1}$. This is similar to the systemic velocity of Sombrero itself 
(1024 kms$^{-1}$; Smith et al. 2000). Furthermore, the radial velocity 
distribution for the bulk of the objects is roughly Gaussian, which is expected for a bound GC system. Thus  we can
conclude that the bulk of the objects with measured velocities 
are probable members of the Sombrero  GC system. The tail of objects with very low and high radial 
velocities in this figure are likely foreground stars and background galaxies respectively. Given the clear gap between the 
bulk of the objects and those with high velocities, we have chosen to restrict our sample to v$_r$ $<$ 2000 kms$^{-1}$
(the highest measured velocity, at $\sim 3800$ kms$^{-1}$, is likely spurious).
We note that the GC candidates show no obvious trend of velocity with either magnitude or colour.

\begin{figure}
\includegraphics[width=88mm]{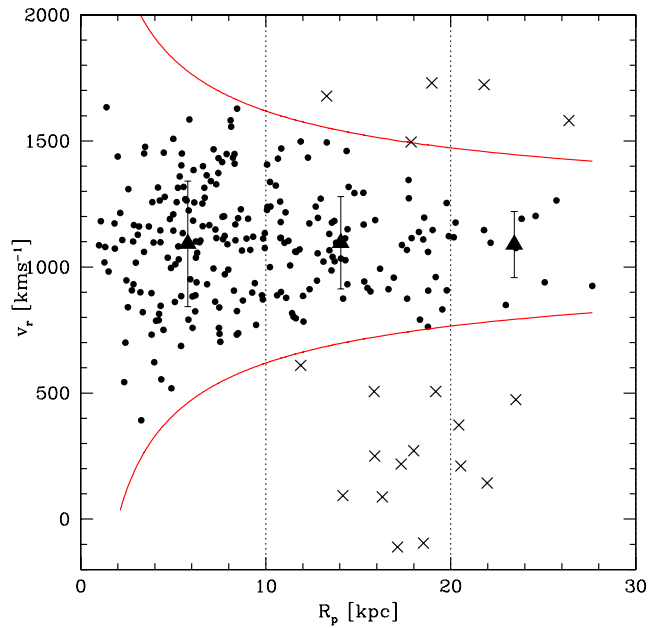}
 \caption{Radial velocities for the Sombrero GC candidates against projected galactocentric distance. 
 The final sample of confirmed GCs are shown by filled circles.
 Excluded objects are
 represented by crosses. The 
 filled triangles indicate the median and dispersion of the radial
 velocities in different radial bins, as marked by the vertical dotted lines.
The pair of solid lines shows a R$^{-1/2}$ dependence to guide the eye.
}
 \label{f:vrgal}
\end{figure}

Figure \ref{f:vrgal} displays the radial velocities versus the projected
galactocentric distance for GC candidates. The bulk of objects have velocities
around 1100  kms$^{-1}$ irrespective of galactocentric distance, although the 
dispersion decreases with radius. Such behaviour is similar to that seen by
Bridges et al. (2007) in their study of the Sombrero GC system. 

We exclude a further 14 objects with low velocities that lie at large 
projected distances (these are likely to be foreground stars). There are also
five objects at large distances with somewhat higher velocities 
than the spread in velocities at that radius. 
These objects may be GCs with anomalous velocities. For the purposes of this 
spectroscopic metallicity analysis we adopt a conservative approach and exclude them from selection. For a kinematic analysis, in which outliers can have a large effect on results, a more 
sophisticated selection process will be required. 

To quantify the velocity 
dispersion we binned our sample into three different radii (0
$<$ R$_{\rm p}$ $\leq$ 10 kpc, 10 $<$ R$_{\rm p}$ $\leq$ 20 kpc and R$_{\rm p}$
$>$ 20 kpc), which was a purely arbitrary choice. 
We measure $\sigma_r$ = 249 km$s^{-1}$ in the first
bin, $\sigma_r$ = 183 km$s^{-1}$ in the second, and $\sigma_r$ = 131
km$s^{-1}$ in the third one. The excluded objects lie at least 2$\sigma$ away from the mean velocity in each bin. 

After excluding the probable foreground stars, background galaxies and high velocity objects we have a
sample of 239 velocity confirmed GCs associated with 
the Sombrero galaxy.

\begin{figure}
\includegraphics[width=88mm]{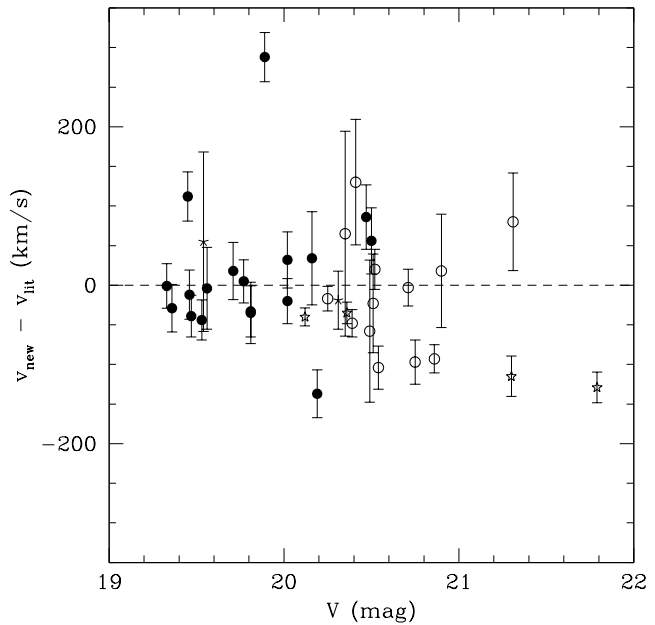}
 \caption{Radial velocity difference between our new measurements and values in
 the literature for the 37 GCs in common. The symbols represent the different
 literature datasets: filled circles (2dF, Bridges et al.~2007); open circles
 (WHT, Bridges et al.~1997); open stars (Keck, Larsen et al.~2002); skeletal
 stars (WIYN, Bridges et al.~2007).
}
 \label{f:vel_lit}
\end{figure}

We have reliable radial velocities for 37 GCs with previously published data. We compare our new velocities
to those from the literature in Fig. \ref{f:vel_lit}, in which the different
studies are coded by symbol type. Nearly all of the repeats are from the 2dF and
WHT studies of Bridges et al.~(1997; 2007). There is no evidence for a zeropoint offset
in the velocity scales, with median differences (in the stated sense of
new$-$old) of $-3$ and $-17$ kms$^{-1}$ for the two studies respectively. In addition,
the distribution of velocities differences generally seems consistent with that
expected from the stated uncertainties, excepting a small number of outliers.
The radial velocities from Larsen et al.~(2002) are systematically higher than
ours ($-78$ in the median), but with only four GCs in common, this conclusion is
not strong.

\subsection{Metallicity Analysis}

In order to derive spectroscopic metallicities for our moderate resolution and moderate signal-to-noise (see Fig. 1) data we adopt a similar 
approach to that of Perrett et al. (2002) and Brodie \& Huchra (1990) who studied the GCs of M31. 
Aiming to measure the metallicities of extragalactic GCs, Brodie \& Huchra
(1990, hereafter BH90) defined a set of calibrations based upon six different
absorption-line indices, which were strongly
sensitive to metallicity. Their metallicity calibrations were tied to the Milky Way GC system and therefore the 
assumption that the GCs are mostly old (i.e. $\geq$ 10 Gyr).

Originally, BH90 measured the metallicity of globular clusters by taking the
weighted mean of the six elemental absorption-line indices, which lead
to estimated metallicities with an accuracy of approximately
15 per cent. However, Perrett et al. (2002) used the same
method employed by BH90 but extended their analysis to 12 absorption-line
indices. Comparing the line indices with published (and
independent) metallicities of GCs in M31, Perrett et al. (2002) found that
the CH (G Band), Mgb and Fe53 line indices were the best metallicity
calibrators. Consequently, they calculated the final metallicities of their
sample by adopting an unweighted mean of these best [Fe/H]-indicators. 

As we are dealing with uncalibrated flux spectra, we have
preferentially used the relatively narrow metallicity indices CH, Mgb and
Fe5270, as they are less sensitive to index calibration errors. In addition, CH
and Mgb constitute some of the strongest features in the integrated spectra of
late-type objects. However, it is worth
emphasizing that due to the different slit positions in the masks, the
DEIMOS spectra cover different wavelengths, and thus it was not
possible to calculate the three indices for all GCs. 

The feature passbands, pseudocontinua
and [Fe/H]-calibrations adopted in this work are exactly as defined in BH90.
The three indices were measured automatically by employing a
modified version of the LECTOR code, which was made available by Alexandre
Vazdekis. The code takes into account the index-band definitions (see
Table \ref{t:index}) and the radial velocities to calculate the line indices and their errors 
accordingly. In Fig. 1 we show the location of the three indices on example spectra. 

\begin{figure}
\includegraphics[width=88mm]{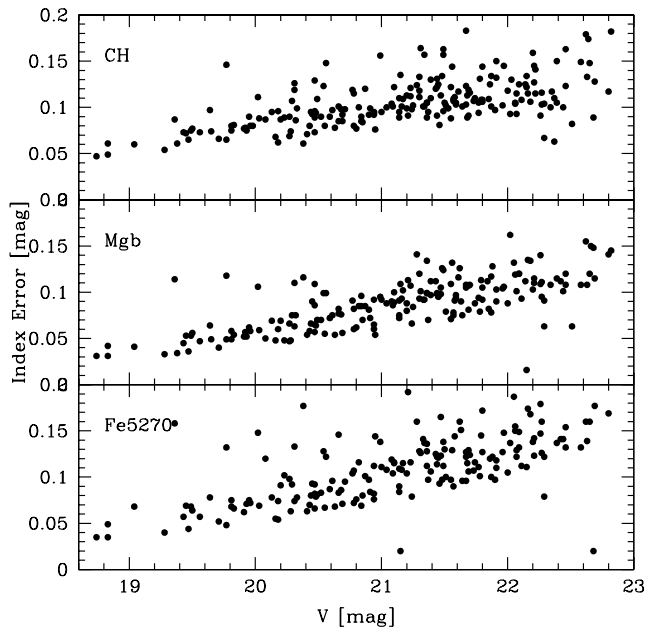}
 \caption{Index error versus GC magnitude. Three indices are shown with CH ({\it top}), Mgb ({\it
 middle}) and Fe5270 ({\it bottom}).}
 \label{f:errorindex}
\end{figure}

In Table B we list the GC magnitudes, colours, raw indices and spectroscopic
metallicities obtained by following the prescription above. The
values given in brackets (when possible) correspond to the estimated
uncertainty on the indices and metallicities. The former corresponds to the
Poisson noise errors, while the latter was estimated by propagating these errors and those associated with the 
metallicity calibrations themselves (see last column of Table
\ref{t:index}) in quadrature.
Note, however, that these uncertainties could be underestimated
since we are not taking into account other possible sources of error.
In Fig. \ref{f:errorindex} we show the Poisson errors as a
function of magnitude for each index. As expected, the uncertainties are
higher for the fainter objects, since they have lower S/N.

\begin{table*}
\begin{minipage}{140mm}
\caption{Definition of BH90 indices and metallicity calibrations.}
\label{t:index}
  \begin{tabular}{@{}lccccl@{}}
  \hline

   Index & Blue & Feature & 
   Red & Width & [Fe/H] calibration  \\
   & $[\rm \AA]$ & $[\rm \AA]$ & $[\rm \AA]$ & $[\rm \AA]$ &   \\

 \hline
CH          & 4268.25-4283.25    & 4283.25-4317.00 &  4320.75-4335.75 & 33.25 &  [Fe/H] = 11.415$\times$(CH) --2.455 ($\sigma$ = 0.26) \\ 
Mgb         & 5144.50-5162.00    & 5162.00-5193.25 &  5193.25-5207.00 & 31.25 &  [Fe/H] = 14.171$\times$(Mgb) --2.216 ($\sigma$ = 0.35)\\ 
Fe5270      & 5235.50-5249.00    & 5248.00-5268.75 &  5288.00-5319.25 & 20.75 &  [Fe/H] = 20.367$\times$(Fe5270) --2.086 ($\sigma$ = 0.33)\\
\hline
\end{tabular}
\end{minipage}
\end{table*}

\begin{figure}
\includegraphics[width=88mm]{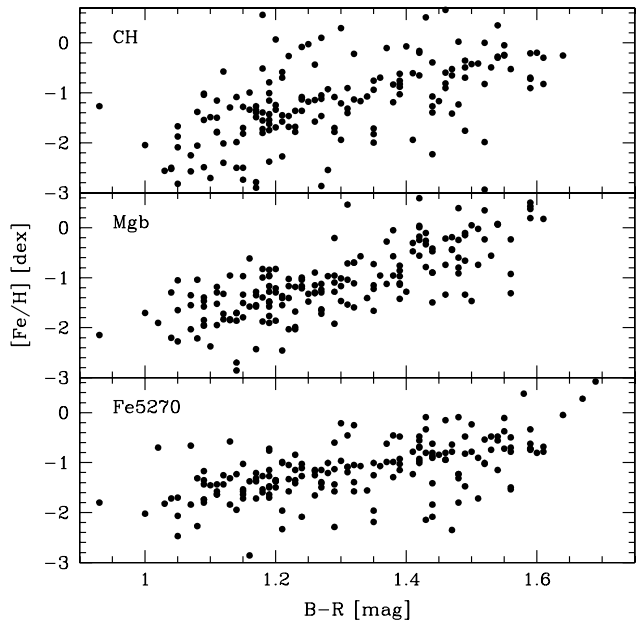}
 \caption{Colour-metallicity relations based on B-R colours and CH ({\it top}), Mgb ({\it middle}) and
 Fe5270 ({\it bottom}) indices.}
 \label{f:fehindex_cor}
\end{figure}

In Fig. \ref{f:fehindex_cor} we plot the derived [Fe/H] metallicity from each individual index against the B--R colour. In old stellar populations, colour is a good proxy for metallicity. This figure shows that all three indices reveal similar trends with respect to B--R colour. The scatter for the 
Mg and Fe5270 indices are similar. The scatter for the CH index is somewhat larger this is because it is a wider index and located at bluer wavelengths for which DEIMOS is less sensitive. We combine the individual 
[Fe/H] values, weighted by their errors, to form a final metallicity of each GC.

For the subsequent metallicity analysis, we have restricted our sample to the brightest objects (i.e. V $\leq$ 22 mag) and those with Poisson errors $<$ 
0.15 dex (see Fig. \ref{f:errorindex}). 
After excluding these lower S/N GCs, 
the remaining 112 GCs have a metallicity range of 
$-$2.2$<$ [Fe/H] $<$ +0.1. 
Interestingly,  Spitler et al.
(2006) found the $B-R$ and $B-V$ colours to give similar metallicity ranges of 
--2.1 $<$ [Fe/H]$_{\rm
B-R}$ $<$ 0.3 and  --2.3 $<$ [Fe/H]$_{\rm B-V}$ $<$ +0.4,
respectively, based on data from the HST/ACS. Hence, within the uncertainties, the
spectroscopic metallicities we derive here are in good agreement with
the photometric ones. 
For example, the mean spectroscopic metallicity of [Fe/H] $\sim$ --1.2 agrees well
with that estimated previously by Bridges et al. (1997, 2007) and Larsen et al.
(2002). Furthermore, the mean difference in [Fe/H] for the four globular clusters in common
with those studied in Larsen et al. (2002) is only --0.14 $\pm$ 0.07 dex (our -
theirs), which is small within the uncertainties.

\subsection{Metallicity Distribution}

In order to test for the presence of bimodality in the GC metallicity distribution
we have employed the KMM algorithm.
The KMM algorithm (see Ashman, Bird \& Zepf
1994 for details) returns the likelihood ratio test statistic and calculates the
probability P of a distribution being unimodal (single Gaussian) or bimodal
(double Gaussian model). It is independent of data binning.

\begin{figure}
\includegraphics[width=88mm]{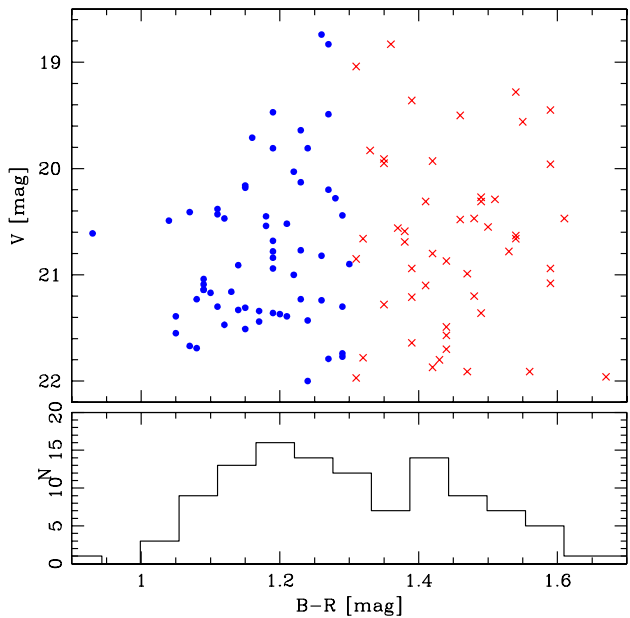}
 \caption{Colour-magnitude diagram ({\it top}) and the colour histogram
 ({\it bottom}) for 112 GCs. The blue points and red crosses
 represent blue and red GC subpopulations cut at B--R = 1.30 respectively.
}
 \label{f:tiltcor}
\end{figure}

For a homoscedastic fit to the $B-R$ colours we find
that the colour distribution for our sample of 239 
spectroscopically-confirmed GCs in the Sombrero galaxy is bimodal at a
confidence level of greater than 99.7 per cent (see Fig. \ref{f:tiltcor}). The blue and
red peaks are at $B-R$ = 1.19 and 1.46 mag, respectively. 
Thus we are sampling a bimodal colour distribution of GCs.

\begin{figure}
\includegraphics[width=88mm]{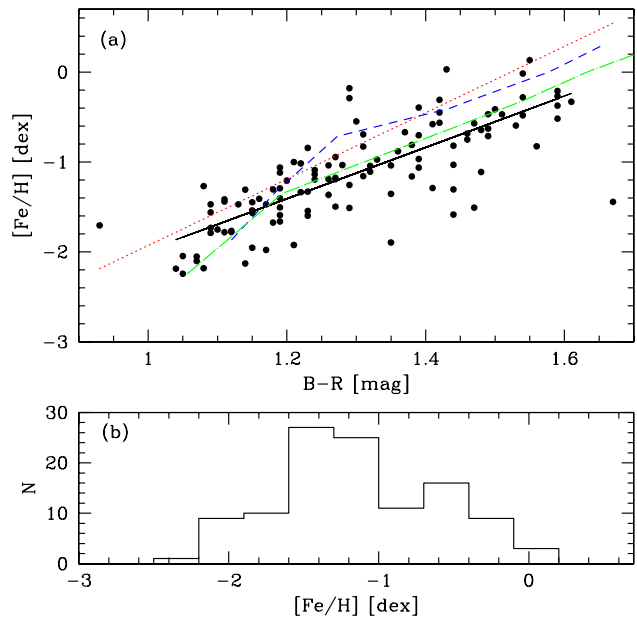}
 \caption{Colour-metallicity relation ({\it top}) and metallicity histogram ({\it bottom}) based on spectroscopic metallicities. The least-squares linear fit is shown by a black solid line. The red dotted line indicates the linear
 colour-metallicity relation found by Barmby et al. (2000) for the GC system of
 M31. The colour-metallicity relations predicted by SSP models are also shown, i.e. Teramo-SpOT (Raimondo et al. 2005; blue short dashed line) and Maraston (1998; green long dashed line).}
 \label{f:metalcor}
\end{figure}

In Fig. \ref{f:metalcor} we show the relationship between GC colour and our derived mean spectroscopic metallicity. The figure shows a good correlation between colour and metallicity with no obvious indication that the relationship is non-linear.
A least-square linear fit to the data gives [Fe/H] =
(--4.83 $\pm$ 0.27) + (2.85 $\pm$ 0.20)$\times$(B-R) with $\sigma$ = 0.33 dex,
for 1 $\leq$ (B-R) $\leq$ 1.65.
The rms dispersion in the relation is comparable to that for the conversion of the raw BH90 indices into metallicity. 
We also show the linear relation of Barmby et al. (2000) who used the BH90
indices for GCs in M31. For comparison, we have also overplotted 
the colour-metallicity relations predicted by the old age simple stellar population (SSP) models of the Teramo-SPoT group (Raimondo
et al. 2005) and Maraston (1998). All provide a reasonable description of 
the data, albeit with small offsets from the least-square best fit. 

For a KMM fit to the spectroscopic metallicities, we find bimodality with a
probability $>$ 90 per cent, where the metal-poor subpopulation is peaked at
[Fe/H] = --1.44 and the metal-rich
one at [Fe/H] = --0.60. These values can be compared to [Fe/H] = --1.38 and -0.49 quoted by Spitler et al. (2006) based on optical colours and [Fe/H] $\sim$ --1.7 and --0.7 from Held et al. (2003) from spectra.

\subsection{Radial Metallicity Profile}

Currently very few galaxies have published radial metallicity profiles for GCs based on spectroscopy. In both the Milky Way (Harris 2001) 
and M31 (Alves-Brito et al. 2009), the GCs follow a similar pattern, i.e. a
strong metallicity gradient in the inner few kpc, which then flattens to a null
gradient in the outer halo. Similar metallicity profiles are inferred for GCs in large elliptical galaxies from their colours (see Forbes et al. 2011 and references therein).  These metallicity profiles suggest an inner region in which gas dissipation has played an important role, and an outer halo region that is dominated by the accretion of small galaxies and their GCs. 

\begin{figure}
\includegraphics[width=88mm]{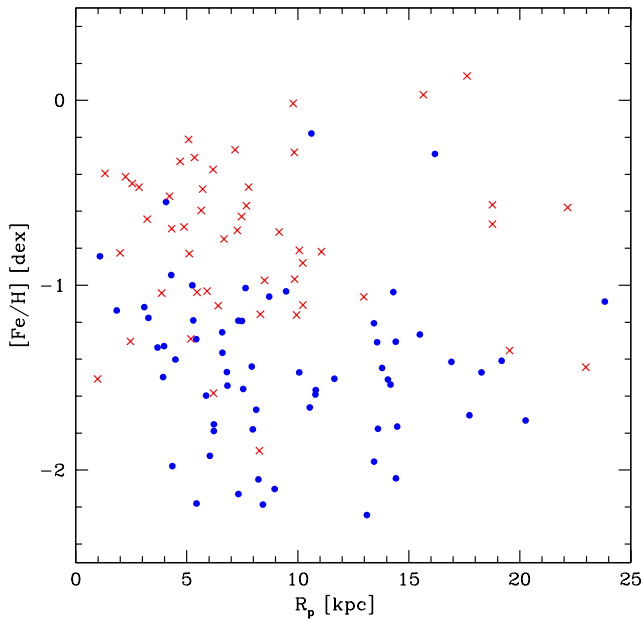}
 \caption{Metallicity as a function of projected galactocentric distance. Symbols are as in Fig. 6.
}
 \label{f:gradiente}
\end{figure}

In Fig. \ref{f:gradiente} we show the radial metallicity profile for the GCs in
the Sombrero galaxy out to 30 kpc. There is no clear indication of metallicity
gradients in either GC subpopulation. However, our data are not uniformly distributed in radius and are quite limited beyond 20 kpc. 
We note that the transition from a strong inner gradient to a null gradient in the massive ellipticals M87 (Harris 2009) and NGC 1407 (Forbes et al. 2011) were seen at around 70 kpc. Given that the luminosity of Sombrero is more comparable to these galaxies than the Milky Way or M31, we might expect the transition radius to occur beyond our field-of-view.

\subsection{Blue Tilt}

It has been shown in several extragalactic GC systems that the blue GC subpopulation 
colour distribution tends to become redder for higher luminosities 
(i.e., a blue tilt). 
While this effect has been found in galaxies of different morphological types
(e.g. Harris et al. 2006; Strader et al. 2006; Mieske et al.
2006; Lee et al. 2008; Harris 2009; Peng et al. 2009; Mieske et al. 2010; Forbes
et al. 2010; Faifer et al. 2011), the Milky Way is a galaxy that do
not reveal a blue tilt to date (e.g., Strader \& Smith 2008). 
According to the scenario proposed by Bailin \& Harris (2009), the
more massive GCs are also more heavy-element enriched. Thus, the fact that the
Milky Way sample includes fewer massive GCs and more lower mass GCs than other
datasets, could account for the lack of tilt in the Milky Way.

On the other hand, Forbes et al. (2010) have recently shown that
there is a blue tilt in the blue GC subpopulation of the Milky
Way analog NGC 5170.
For the Sombrero galaxy, Spitler et al. (2006) have shown photometrically that its
blue GCs reveal a blue tilt not only at the brightest magnitudes (highest mass) 
but that it may also extend to relatively low mass GCs too. 
The blue tilt is generally assumed to be a 
mass-metallicity trend that is due to a self-enrichment process (Strader \& Smith 2008; Bailin \& Harris 2009). 

\begin{figure}
\includegraphics[width=88mm]{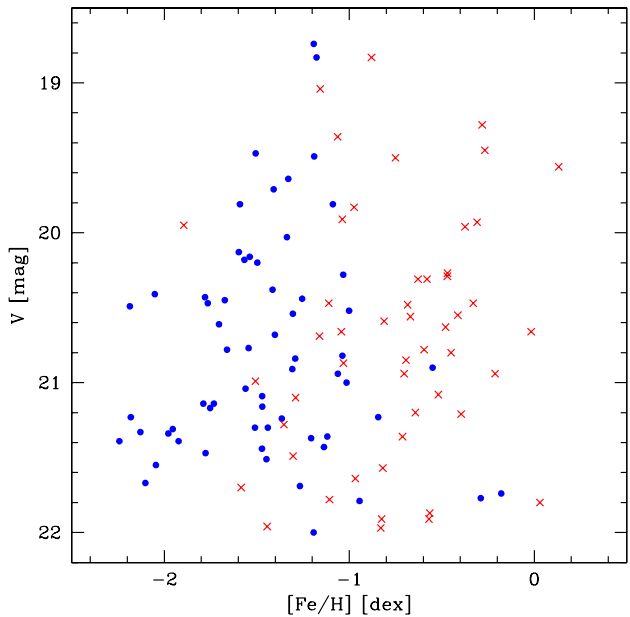}
 \caption{Metallicity-magnitude diagram.  Symbols are as in
 Fig. \ref{f:tiltcor}.}
 \label{f:tiltfeh}
\end{figure}

We present in Fig. \ref{f:tiltfeh} the spectroscopic metallicity-magnitude
diagram for Sombrero GCs. It shows that the brightest GCs have magnitudes comparable to Omega Cen which would have V $\sim$ 19.5 if located at the distance of Sombrero. 
Although the brightest blue GCs are more metal-rich than average, 
our number statistics are too low to make any definitive statements about a spectroscopic `blue tilt' in the Sombrero galaxy GC system. In the self-enrichment models mentioned above the tilt is present for GCs with masses above a million solar masses or V $\le$ 19.5.  

\section{Summary and conclusions}

Globular clusters are widely recognised as extremely useful tracers of the
formation and evolution of their host galaxies. Here we present Keck spectra for over 200 GCs with radial velocities that confirm their 
association  
with the Sombrero galaxy. For many of the GCs the spectra are of sufficient quality to derive Brodie \& Huchra (1990) 
style metallicities from several indices. 

We find that the GCs span a broad metallicity range from [Fe/H] = --2.2 to
+0.1. This is comparable to previous estimates based on photometric metallicities. 
For a restricted high quality subsample of 112 GCs, we find a good correlation
between individual GC metallicity and B--R colour. The resulting spectroscopic metallicity distribution is clearly bimodal with peaks at 
[Fe/H] $\sim$ --1.4 and --0.6, 
with a statistical probability of $>$ 90 per cent. Thus despite claims that colour bimodality may not reflect metallicity bimodality in extragalactic GC 
systems,  it would appear that the GC system of the Sombrero galaxy joins that of some other giant elliptical galaxies in revealing spectroscopic bimodality. 
Furthermore we find a transformation 
between optical colour and metallicity that does not require a non-linear relation.
 
We also investigate the GC radial metallicity profile out to 30 kpc and a spectroscopic version of the colour-magnitude diagram, but our small number statistics prevent any strong statements about trends in the GC subpopulations.

\section*{Acknowledgments}
AAB acknowledges CNPq (PDE, 200227/2008-4) and FONDECYT (3100013) for financial
support. Likewise, GKTH, DF and LP thank ARC for financial support. JD
acknowledges support from NSF grants AST-0917706 and AST-0909237. We are
greatful to Caroline Foster and Bill Harris for useful discussions and
suggestions. We thank an anonymous referee for useful comments and
suggestions. Support for this work was provided by award 1310512, issued by JPL/Caltech.
The analysis pipeline used to reduce the DEIMOS data was developed at
UC Berkeley with support from NSF grant AST-0071048. This work was supported by the National Science Foundation
through grant AST-0808099. We thank Steve Zepf for his help supplying an object list of globular candidates,
and Arunav Kundu for help with astrometry.

\appendix

\section{Positions and velocities}


\begin{table*}
 \centering
 \begin{minipage}{140mm}
  \caption{IDs, positions and velocities for the sample.}
\begin{tabular}{@{}lcccccccccc@{}}
  \hline
   ID & Alt ID & Mask & R.A. & Dec. & X & Y & R$_{\rm p}$ & V$_{\rm r}$ &
   $\sigma$V$_{\rm r}$ \\ 
    &   &   &  (J2000) & (J2000) & (arcmin) & (arcmin) & (arcmin) & kms$^{-1}$ & kms$^{-1}$  \\ 
      (1) & (2) & (3) & (4) & (5) & (6) & (7) & (8) & (9) & (10) \\ 
      
 \hline
RZ1157 & -- & 1 & 12 40 36.93 & -11 41 17.55 & 9.224 & -3.907 & 10.020 & 1581 & 49 \\
RZ1191 & -- & 1 & 12 40 36.35 & -11 40 58.07 & 9.082 & -3.582 & 9.763 & 1264 & 8 \\
RZ1192 & -- & 1 & 12 40 36.34 & -11 39 33.72 & 9.079 & -2.177 & 9.336 & 1202 & 10 \\
RZ1210 & -- & 1 & 12 40 36.03 & -11 38 19.15 & 9.001 & -0.933 & 9.049 & 1191 & 117 \\
RZ1255 & -- & 1 & 12 40 34.76 & -11 38 10.23 & 8.689 & -0.785 & 8.724 & 849 & 8 \\
\hline
\end{tabular}
\end{minipage}
\begin{minipage}{.88\hsize}
 
 \end{minipage}		
\end{table*}

\begin{table*}
 \centering
 \begin{minipage}{140mm}
 \contcaption{IDs, positions and velocities for the sample.}
  \begin{tabular}{@{}lcccccccccc@{}}
  \hline
  ID & Alt ID & Mask & R.A. & Dec. & X & Y & R$_{\rm p}$ & V$_{\rm r}$ &
   $\sigma$V$_{\rm r}$ \\ 
    &   &   &  (J2000) & (J2000) & (arcmin) & (arcmin) & (arcmin) & kms$^{-1}$ & kms$^{-1}$  \\ 
      (1) & (2) & (3) & (4) & (5) & (6) & (7) & (8) & (9) & (10) \\ 
 \hline
RZ2475 & --   & 4 & 12 40 07.68 & -11 46 04.80 & 2.031 & -8.697 & 8.931 & 1075 & 9 \\
RZ2491 & S059 & 3 & 12 40 07.49 & -11 35 51.47 & 1.985 & 1.526 & 2.503 & 1162 & 12 \\
RZ2493 & S180 & 3 & 12 40 07.45 & -11 36 13.12 & 1.975 & 1.165 & 2.293 & 897 & 32 \\
RZ2502 & S030 & 1 & 12 40 07.26 & -11 36 44.65 & 1.928 & 0.639 & 2.031 & 1359 & 8 \\
RZ2503 & --   & 2 & 12 40 07.25 & -11 40 25.42 & 1.924 & -3.040 & 3.598 & 770 & 10 \\
\hline
\end{tabular}
\end{minipage}
\end{table*}

\begin{table*}
 \centering
 \begin{minipage}{140mm}
  \contcaption{IDs, positions and velocities for the sample.}
  \begin{tabular}{@{}lcccccccccc@{}}
  \hline
  ID & Alt ID & Mask & R.A. & Dec. & X & Y & R$_{\rm p}$ & V$_{\rm r}$ &
   $\sigma$V$_{\rm r}$ \\ 
    &   &   &  (J2000) & (J2000) & (arcmin) & (arcmin) & (arcmin) & kms$^{-1}$ & kms$^{-1}$  \\ 
      (1) & (2) & (3) & (4) & (5) & (6) & (7) & (8) & (9) & (10) \\ 
 \hline
RZ3222 & S026 & 1 & 12 39 51.88 & -11 36 27.15 & -1.854 & 0.930 & 2.075 & 1450 & 8 \\RZ3226 & S112 & 4 & 12 39 51.80 & -11 39 24.71 & -1.876 & -2.028 & 2.763 & 1149 & 8 \\RZ3236 & S420 & 2 & 12 39 51.62 & -11 36 07.66 & -1.920 & 1.256 & 2.294 & 759 & 10 \\RZ3244 & S084 & 2 & 12 39 51.51 & -11 36 28.83 & -1.946 & 0.902 & 2.145 & 1269 & 8 \\RZ3261 & S433 & 2 & 12 39 51.24 & -11 38 17.57 & -2.013 & -0.909 & 2.209 & 791 & 21 \\
\hline
\end{tabular}
\end{minipage}
\end{table*}

\begin{table*}
 \centering
 \begin{minipage}{140mm}
  \contcaption{IDs, positions and velocities for the sample.}
  \begin{tabular}{@{}lcccccccccc@{}}
  \hline
  ID & Alt ID & Mask & R.A. & Dec. & X & Y & R$_{\rm p}$ & V$_{\rm r}$ &
   $\sigma$V$_{\rm r}$ \\ 
    &   &   &  (J2000) & (J2000) & (arcmin) & (arcmin) & (arcmin) & kms$^{-1}$ & kms$^{-1}$  \\ 
      (1) & (2) & (3) & (4) & (5) & (6) & (7) & (8) & (9) & (10) \\ 
 \hline
S003 & -- & 4 & 12 40 00.09 & -11 36 09.13 & 0.164 & 1.231 & 1.242 & 392 & 8 \\S004 & -- & 2 & 12 40 14.59 & -11 38 27.99 & 3.729 & -1.083 & 3.883 & 999 & 8 \\S007 & -- & 1 & 12 39 48.62 & -11 35 40.63 & -2.657 & 1.706 & 3.158 & 1410 & 8 \\S017 & -- & 4 & 12 40 00.05 & -11 35 52.92 & 0.155 & 1.501 & 1.509 & 622 & 10 \\S027 & -- & 3 & 12 40 00.29 & -11 37 54.80 & 0.212 & -0.530 & 0.571 & 982 & 8 \\
\hline
\end{tabular}
\end{minipage}
\end{table*}

\begin{table*}
 \centering
 \begin{minipage}{140mm}
\contcaption{IDs, positions and velocities for the sample.}
  \begin{tabular}{@{}lcccccccccc@{}}
  \hline
  ID & Alt ID & Mask & R.A. & Dec. & X & Y & R$_{\rm p}$ & V$_{\rm r}$ &
   $\sigma$V$_{\rm r}$ \\ 
    &   &   &  (J2000) & (J2000) & (arcmin) & (arcmin) & (arcmin) & kms$^{-1}$ & kms$^{-1}$  \\ 
      (1) & (2) & (3) & (4) & (5) & (6) & (7) & (8) & (9) & (10) \\ 
 \hline
S358 & -- & 2 & 12 40 11.51 & -11 39 18.11 & 2.974 & -1.918 & 3.539 & 1120 & 12 \\S363 & -- & 1 & 12 40 03.37 & -11 36 31.16 & 0.971 & 0.864 & 1.300 & 1450 & 34 \\S368 & -- & 1 & 12 40 05.39 & -11 37 59.00 & 1.466 & -0.600 & 1.584 & 1315 & 28 \\S369 & -- & 4 & 12 39 59.75 & -11 37 54.27 & 0.079 & -0.521 & 0.527 & 1635 & 14 \\S370 & -- & 1 & 12 39 51.27 & -11 34 55.48 & -2.006 & 2.459 & 3.173 & 1169 & 8 \\S376 & -- & 3 & 12 40 10.47 & -11 36 36.33 & 2.717 & 0.778 & 2.826 & 1433 & 18 \\S384 & -- & 4 & 12 39 58.84 & -11 38 11.39 & -0.144 & -0.806 & 0.819 & 990 & 22 \\S388 & -- & 1 & 12 40 07.73 & -11 38 14.33 & 2.043 & -0.855 & 2.214 & 1224 & 15 \\S393 & -- & 3 & 12 39 48.68 & -11 35 39.54 & -2.642 & 1.724 & 3.155 & 1448 & 20 \\S435 & -- & 2 & 12 39 47.96 & -11 36 58.39 & -2.820 & 0.410 & 2.849 & 735 & 17 \\S438 & -- & 3 & 12 40 13.73 & -11 39 54.90 & 3.520 & -2.531 & 4.336 & 817 & 21 \\S449 & -- & 4 & 12 39 55.85 & -11 39 58.24 & -0.879 & -2.587 & 2.733 & 896 & 24 \\S490 & -- & 2 & 12 40 15.77 & -11 39 04.91 & 4.021 & -1.698 & 4.365 & 803 & 11 \\S502 & -- & 3 & 12 40 10.77 & -11 38 18.93 & 2.791 & -0.932 & 2.942 & 1201 & 33 \\S529 & -- & 1 & 12 40 03.53 & -11 37 50.57 & 1.010 & -0.459 & 1.110 & 2285 & 36 \\S600 & -- & 2 & 12 39 41.37 & -11 34 53.43 & -4.439 & 2.493 & 5.091 & 1323 & 8 \\S605 & -- & 2 & 12 39 58.33 & -11 38 41.53 & -0.269 & -1.309 & 1.336 & 1066 & 24 \\S626 & -- & 4 & 12 39 54.17 & -11 38 13.60 & -1.293 & -0.843 & 1.544 & 1691 & 19 \\
\hline
\end{tabular}
\end{minipage}
\end{table*}

\section{Magnitudes, colours, indices and metallicities}

In Table B1 we present magnitudes, colours, indices and metallicities for the globular cluster candidates. The table is ordered by ID.

\begin{table*}
\centering
\setlength\tabcolsep{4pt}
\begin{minipage}{180mm}
\caption{IDs, magnitudes, colours, indices and metallicites for the GC candidate sample.}
  \begin{tabular}{@{}llccccccccc@{}}
  \hline
 
   ID & Alt ID & V & B-R & CH & [Fe/H]$_{\rm CH}$ & Mgb & [Fe/H]$_{\rm Mgb}$ & Fe5270 & [Fe/H]$_{\rm 5270}$ & [Fe/H] \\ 
      (1) & (2) & (3) & (4) & (5) & (6) & (7) & (8) & (9) & (10) & (11) \\ 
\hline 

RZ1191  & -- &  20.37  &  1.22  &   0.170 (0.007)   &   -0.515 (0.079)   &   0.064 (0.005)   &   -1.309 (0.073)   &   0.053 (0.004)   &   -1.007  (0.085)   &   -0.937 (0.045)     \\RZ1192  & -- &  21.20  &  1.05  &   0.032 (0.009)   &   -2.090 (0.113)   &   0.040 (0.006)   &   -1.649 (0.080)   &    --        &   --         &    --          \\RZ1210  & -- &  22.66  &  1.09  &   0.047 ( -- )   &   -1.918 ( -- )   &   0.026 (0.010)   &   -1.848 (0.149)   &   0.032 (0.011)   &   -1.434  (0.217)   &   -1.917 (--)     \\RZ1255  & -- &  19.81  &  1.24  &   0.116 (0.006)   &   -1.131 (0.075)   &   0.086 (0.003)   &   -0.997 (0.049)   &   0.048 (0.003)   &   -1.108  (0.074)   &   -1.089 (0.038)     \\RZ1271  & -- &  21.96  &  1.67  &  -0.320 (0.009)   &   -6.108 (0.102)   &   0.229 (0.007)   &   1.029 (0.103)   &   0.116 (0.006)   &    0.277  (0.127)   &   -1.444 (0.064)     \\RZ1290  & -- &  22.35  &  0.97  &  -0.014 (0.013)   &   -2.615 (0.149)   &   -0.055 (0.017)   &   -2.995 (0.237)   &  -0.031 (0.011)   &   -2.717  (0.220)   &   -2.801 (0.118)     \\
\hline					    
\end{tabular}				    
\end{minipage}	
\end{table*}

\begin{table*}
 \centering
 \setlength\tabcolsep{4pt}
 \begin{minipage}{180mm}
\contcaption{IDs, magnitudes, colours, indices and metallicites for the sample.}
  \begin{tabular}{@{}llccccccccc@{}}
  \hline
   ID & Alt ID & V & B-R & CH & [Fe/H]$_{\rm CH}$ & Mgb & [Fe/H]$_{\rm Mgb}$ & Fe5270 & [Fe/H]$_{\rm 5270}$ & [Fe/H] \\ 
      (1) & (2) & (3) & (4) & (5) & (6) & (7) & (8) & (9) & (10) & (11) \\ 
\hline  

RZ2475  & -- &  20.02  &  1.19  &   0.090 (0.010)  &    -1.428 (0.111)   &   0.052 (0.008)  &   -1.479 (0.106)  &   --   --     &    --    &    --            \\RZ2491  &  S059  &  20.44  &  1.29  &   0.120 (0.008)  &   -1.085 (0.095)  &   0.066  (0.005)  &   -1.281  (0.066)  &   0.032 (0.004)  &   -1.434 (0.080)  &    -1.255  (0.046)  \\RZ2493  &  S180  &  21.39  &  1.21  &   0.016 (0.011)  &   -2.272 (0.130)   &   -0.017  (0.007)  &   -2.457 (0.097)   &   0.053 (0.005)  &   -1.007  (0.106)  &    -1.923  (0.064)  \\RZ2502  &  S030  &  19.93  &  1.42  &   0.198 (0.006)  &   -0.195 (0.075)   &   0.139 (0.004)   &   -0.246 (0.052)   &   0.079 (0.003)   &   -0.477   (0.071)  &   -0.309 (0.038)    \\RZ2503  & -- &  20.28  &  1.28  &   0.134 (0.006)  &   -0.925 (0.074)  &  0.088  (0.003)   &   -0.969 (0.047)   &   0.043 (0.003)   &   -1.210 (0.063)  &     -1.033 (0.036)     \\RZ2532  &  S177  &  21.48  &  1.24  &   0.291 (0.012)  &    0.867 (0.134)  &   --       &   --     &   0.040 (0.005)   &   -1.271 (0.096)  &     --       \\
\hline					    
\end{tabular}				    
\end{minipage}				    
\end{table*}				    

\begin{table*}
 \centering
 \begin{minipage}{180mm}
 \setlength\tabcolsep{4pt}
\contcaption{IDs, magnitudes, colours, indices and metallicites for the sample.}
  \begin{tabular}{@{}llccccccccc@{}}
  \hline
   ID & Alt ID & V & B-R & CH & [Fe/H]$_{\rm CH}$ & Mgb & [Fe/H]$_{\rm Mgb}$ & Fe5270 & [Fe/H]$_{\rm 5270}$ & [Fe/H] \\ 
      (1) & (2) & (3) & (4) & (5) & (6) & (7) & (8) & (9) & (10) & (11) \\ 
\hline  

RZ3261  &  S433  &  22.46  &  1.20  &   0.019 (0.011)  &   -2.238 (0.122)  &   0.025 (0.008)   &   -1.862 (0.108)   &   0.029 (0.006)   &   -1.495 (0.132)  &     -1.856 (0.070)  \\RZ3265  &  S028  &  19.96  &  1.59  &   0.154 (0.007)  &   -0.697 (0.080)  &   0.183  (0.004)  &   0.377  (0.058)  &   0.072 (0.004)  &   -0.620 (0.072)  &    -0.374  (0.040)     \\RZ3274  &  S073  &  20.69  &  1.21  &   0.077 (0.008)  &   -1.576 (0.092)  &   0.047  (--  )  &   -1.550  (--  )  &   -0.012 (--  )  &   -2.330 (--  )  &    -1.940  ( --)  \\RZ3283  &  S188  &  21.44  &  1.18  &   0.264 (0.008)  &    0.559 (0.091)   &    --     &   --       &   0.027 (0.006)   &   -1.536  (0.114)   &    --          \\RZ3319  &  S401  &  22.42  &  1.13  &   0.056 (-- )   &   -1.816 (-- )   &   0.089 (0.008)   &   -0.955 (0.111)   &   0.074 (0.007)   &   -0.579 (0.141)  &     -1.813 (--)  \\RZ3320  &  S013  &  19.50  &  1.46  &   0.163 (0.007)  &   -0.594 (0.077)  &   0.104  (0.004)  &   -0.742  (0.056)  &   0.056 (0.003)  &   -0.945 (0.064)  &    -0.750  (0.038)    \\
\hline					    
\end{tabular}				    
\end{minipage}				    
\end{table*}				    

\begin{table*}
 \centering
 \setlength\tabcolsep{4pt}
 \begin{minipage}{180mm}
\contcaption{IDs, magnitudes, colours, indices and metallicites for the sample.}
  \begin{tabular}{@{}llccccccccc@{}}
  \hline
   ID & Alt ID & V & B-R & CH & [Fe/H]$_{\rm CH}$ & Mgb & [Fe/H]$_{\rm Mgb}$ & Fe5270 & [Fe/H]$_{\rm 5270}$ & [Fe/H] \\ 
      (1) & (2) & (3) & (4) & (5) & (6) & (7) & (8) & (9) & (10) & (11) \\ 
\hline  

S027  & -- &  19.92  &  1.25  &   0.112 (0.007)  &   -1.177 (0.078)  &   0.063  (0.004)  &   -1.323  (0.057)  &   --      &       --     &     --           \\S032  & -- &  19.95  &  1.35  &   0.040 (0.008)  &   -1.998 (0.090)  &   0.039  (0.004)  &   -1.663  (0.062)  &   0.006 (0.004)  &   -1.964 (0.075)  &    -1.895  (0.044)     \\S034  & -- &  20.03  &  1.22  &   0.068 (0.008)  &   -1.679 (0.087)   &   0.074 (0.004)   &   -1.167 (0.059)   &   0.051 (0.003)   &   -1.047   (0.069)  &   -1.337 (0.042)     \\S035  & -- &  20.18  &  1.55  &   0.193 (0.005)  &   -0.252 (0.062)  &   --       &      --    &   0.084 (0.003)   &   -0.375 (0.054)  &     --           \\S037  & -- &  20.08  &  1.17  &   0.104 (0.007)  &   -1.268 (0.086)   &   0.061 (0.003)   &   -1.352 (0.050)   &   --     &     --       &     --         \\S039  & -- &  20.20  &  1.27  &   0.087 (0.008)  &    -1.462 (0.087)   &   0.040 (0.005)  &   -1.649 (0.069)  &   0.033 (0.004)  &   -1.414 (0.091)  &   -1.497  (0.047)  \\
\hline					    
\end{tabular}				    
\end{minipage}				    
\end{table*}				    

\begin{table*}
 \centering
 \setlength\tabcolsep{4pt}
 \begin{minipage}{180mm}
\contcaption{IDs, magnitudes, colours, indices and metallicites for the sample.}
  \begin{tabular}{@{}llccccccccc@{}}
  \hline
   ID & Alt ID & V & B-R & CH & [Fe/H]$_{\rm CH}$ & Mgb & [Fe/H]$_{\rm Mgb}$ & Fe5270 & [Fe/H]$_{\rm 5270}$ & [Fe/H] \\ 
      (1) & (2) & (3) & (4) & (5) & (6) & (7) & (8) & (9) & (10) & (11) \\ 
\hline  

S370  & -- &  22.15  &  1.16  &   0.079 (0.012)  &   -1.553 (0.134)   &   --       &      --      &  -0.038 (0.010)   &   -2.860  (0.207)   &    --          \\S376  & -- &  22.21  &  1.32  &  -0.126 (0.013)  &   -3.893 (0.145)  &   0.044  (0.008)  &   -1.592  (0.110)  &   0.090 (0.006)  &   -0.253 (0.123)  &    -2.039  (0.073)  \\S388  & -- &  22.33  &  1.15  &  -0.071 ( -- )  &   -3.265 ( -- )   &   0.060 (0.010)   &   -1.366 (0.145)   &   0.118 (0.008)   &   0.317   (0.170)   &   -3.256 (--)     \\S393  & -- &  22.29  &  1.26  &   0.132 (0.006)  &   -0.948 (0.067)  &   0.075  (0.004)  &   -1.153  (0.063)  &   0.041 (0.004)  &   -1.251 (0.079)  &    -1.124  (0.040)  \\S435  & -- &  22.34  &  1.14  &   0.197 (0.009)  &   -0.206 (0.100)  &   0.036 (0.008)   &   -1.706 (0.108)   &   --      &   --      &     --       \\S438  & -- &  22.46  &  1.32  &   0.183 (0.014)  &   -0.366 (0.163)  &   0.107  (0.008)  &   -0.700  (0.120)  &   0.051 (0.008)  &   -1.047 (0.154)  &    -0.697  (0.084)  \\
\hline					    
\end{tabular}				    
\end{minipage}				    
\end{table*}

\end{document}